**Mergers, Radio Jets, and Quenching Star-Formation in Massive Galaxies: Quantifying their Synchronized Cosmic Evolution & Assessing the Energetics**

Timothy M. Heckman, Namrata Roy[1]

Philip N. Best, Rohit Kondapally[2]

1. *The William H. Miller III Department of Physics and Astronomy, The Johns Hopkins University, Baltimore, MD 21218*
2. *Institute for Astronomy, University of Edinburgh, Royal Observatory, Blackford Hill, Edinburgh, EH9 3HJ, UK*

ABSTRACT

The existence of a population of massive quiescent galaxies with little to no star-formation poses a challenge to our understanding of galaxy evolution. The physical process that quenched the star formation in these galaxies is debated, but the most popular possibility is that feedback from supermassive black holes lifts or heats the gas that would otherwise be used to form stars. In this paper, we evaluate this idea in two ways. First, we compare the cumulative growth in the cosmic inventory of the total stellar mass in quiescent galaxies to the corresponding growth in the amount of kinetic energy carried by radio jets. We find that these two inventories are remarkably well-synchronized, with about 50% of the total amounts being created in the epoch from $z \approx 1$ to 2. We also show that these agree extremely well with the corresponding growth in the cumulative number of major mergers that result in massive ($> 10^{11}$ $M_\odot$) galaxies. We therefore argue that major mergers trigger the radio jets and also transform the galaxies from disks to spheroids. Second, we evaluate the total amount of kinetic energy delivered by jets and compare it to the baryonic binding energy of the galaxies. We find the jet kinetic energy is more than sufficient to quench star-formation, and the quenching process should be more effective in more massive galaxies. We show that these results are quantitatively consistent with recent measurements of the Sunyaev-Zel'dovich effect seen in massive galaxies at $z \approx 1$.

Corresponding author: Timothy Heckman
theckma1@jhu.edu

1. INTRODUCTION

One of the most fundamental properties of galaxies is the existence of populations of star-forming and quiescent galaxies. The former shows a relatively tight correlation between star-formation rate (SFR) and stellar mass ($M_*$), forming what is known as the star-forming main sequence. This relationship evolves strongly towards higher SFR/$M_*$ with redshift (e.g. Popesso et al 2023). In contrast, the population of quiescent galaxies exhibits very little or no detectable star-formation, with mass-doubling times ($M_*$/SFR) much greater than the age of the universe (Weaver et al. 2023, Davidzon et al. 2018). The relative numbers of quiescent vs. star forming galaxies increases as function of $M_*$ (e.g. Weaver et al. 2023). The processes that produce quiescent massive galaxies remain unclear, but feedback from accreting supermassive black holes (Active Galactic Nuclei = AGN) is a leading candidate. The effects of this feedback would be to lift and/or heat gas in the galaxy halo and thereby quench future star-formation (e.g. Fabian 2012; Somerville & Davé 2015; Naab & Ostriker 2017; Crain & van de Voort 2023).

The evidence that massive galaxies are quenched by AGN feedback is tantalizing but inconclusive. Observations of AGN-driven winds and jets are plentiful, but their overall impact on galaxy evolution remains uncertain. Recently Heckman & Best (2023 – HB23) computed the cosmic evolution of the global rate of the generation of kinetic energy by AGN jets and winds. For jets, they used the empirical calibration of jet kinetic power as a function of the radio synchrotron luminosity and then integrated over the AGN radio luminosity function from z ≈ 0.1 to 4. Similar analyses have been undertaken by Kondapally et al (2023 - K23) and by Smolĉiĉ et al (2017 – S17).

For AGN winds, HB23 used the estimated kinetic power of outflowing ionized and molecular gas as a function of AGN bolometric luminosity, based on the extensive existing literature. There were several key findings that motivate the current paper. First, they found that jets have most-likely delivered significantly more kinetic energy than winds. Second, they found that the peak in the energy injection rate from jets occurred significantly later than from winds: z ≈ 0.5 to 2 vs. ≈ 1 to 3, respectively. Third, they found that the stellar-to-black-hole mass ratio at which the galaxy population transitions from being mostly star-forming to mostly quenched galaxies, aligns with the value at which feedback switches from being primarily from massive stars to primarily from AGN.

In this paper, we will expand upon the second two items. First, we will compare the evolution of the cumulative deposition of kinetic energy by AGN jets to the cumulative growth in the stellar mass in quiescent galaxies. We will show that these are extremely similar, with roughly half of both quantities being produced during the range z = 1 to 2. Second, we will evaluate the amount of kinetic energy injected by jets per unit stellar mass as a function of stellar mass, and compare this to the baryonic binding energy of the gas to ask if this is sufficient to quench star-formation in the galaxy.

Throughout this paper we adopt the values $H_0$ = 69.6 km s$^{-1}$ Mpc$^{-1}$, $\Omega_M$ = 0.286 and $\Omega_\Lambda$ = 0.714 (Wright 2006).

2. ANALYSIS
  2.1 The Cumulative Growth in the Mass of Quiescent Galaxies

Over the past few years there have been several studies that have measured the stellar mass function for quiescent galaxies at various redshifts (e.g. Weigel et al. 2016; McLeod et al. 2021 – M21; Noirot et al. 2022; Weaver et al. 2023 – W23).

To compute the cumulative growth in the mass in quiescent galaxies in a consistent way across the broadest range in redshift, we use the stellar mass functions of quenched galaxies from M21 and W23[1].

The results of this exercise are shown in Figure 1, showing the cumulative mass relative to the mass at z = 0.35 (the lowest redshift bin in W23). While the formal uncertainties are small in W23 and M21 (< 0.05 dex), we believe the comparison between these two independent analyses is a better indicator of the uncertainties. We note that these uncertainties include the effects of Poisson noise, cosmic variance, and SED fitting. They do not include systematic uncertainties like the nature of the stellar initial mass function (IMF). All the data in Figure 1 are based on the same standard Chabrier (2003) IMF, so this choice should not impact the evolution of the normalized cumulative mass. The average disagreement between the two such calculations is about 0.1 dex.

### 2.2 The Cumulative Growth in the Kinetic Energy Injected by Radio Jets

For convenience, we briefly summarize here the methodology used by HB23 and K23 to compute the evolution of the rate of energy injection by jets.

This begins with converting the radio synchrotron luminosity at a rest frequency of 1.4 GHz ($L_{1.4GHz}$) into a jet mechanical power ($P_{jet}$). Both papers use the extensive set of data on radio-jet-inflated X-ray cavities at low-z that has been used to compute the cavity energetics (e.g Bırzan et al. 2004, 2008; Dunn et al. 2005; Rafferty et al. 2006; Cavagnolo et al. 2010). This assumes the injected jet energy is $E_{cav}$ = 4 PΔV, where the cavities occupy a volume of ΔV in a region with pressure P. This is converted into a cavity power ($P_{cav}$) by dividing $E_{cav}$ by the cavity buoyancy or sound-crossing timescale (e.g. Churazov et al. 2002). This power requirement for cavity inflation is then provided by the jet kinetic luminosity ($P_{cav}$ = $P_{jet}$).

HB23, K23, and S17 used two different fits to the resulting correlation between $L_{1.4GHz}$ and $P_{cav}$. K23 used a purely empirical approach where the slope and normalization are fit to the data from the literature:

1. $P_{cav}$ = $P_{jet}$ = 1.3 × $10^{38}$ ($L_{1.4GHz}$/$10^{26}$ W Hz$^{-1}$)$^{0.68}$ W

Willott et al. (1999) derive an alternative scaling based on how the ratio of the radio lobe minimum-energy divided by the hot spot dynamical time scales with the synchrotron luminosity. This defines a slope for the relation, but the normalization of the relation was uncertain. HB23 used the literature data on X-ray cavities to determine the normalization:

---

[1] McLeod et al. only compute the mass function up to z = 3.25 and the uncertainties in the Weaver et al. values are relatively large above this redshift. However, these papers show a negligible amount of stellar mass was accumulated in quiescent galaxies at z > 3.25.

2. $P_{cav} = P_{jet} = 1.6 \times 10^{38} (L_{1.4GHz}/10^{26}\ W\ Hz^{-1})^{0.84}$ W

The HB23 formulation leads to larger values of $P_{jet}$ for $L_{1.4GHz} > 10^{26}$ W Hz$^{-1}$ and smaller values at lower radio luminosities. K17 used the Willot et al. formalism and showed results based on a range of normalizations.

These formulae are then applied to the measured radio luminosity functions at various redshifts to calculate the integrated amount of jet kinetic energy injection per co-moving volume element per Gyr at each epoch. An interpolation between epochs leads to the rate of jet energy injection per unit volume and unit time as a function of redshift. HB23 used the empirical models of Yuan et al. (2017) for the evolution of the radio luminosity function. K23 used the radio luminosity functions from Kondapally et al. (2022) based on LoTSS, and S17 used the VLA-COSMOS 3GHz Large Project data.

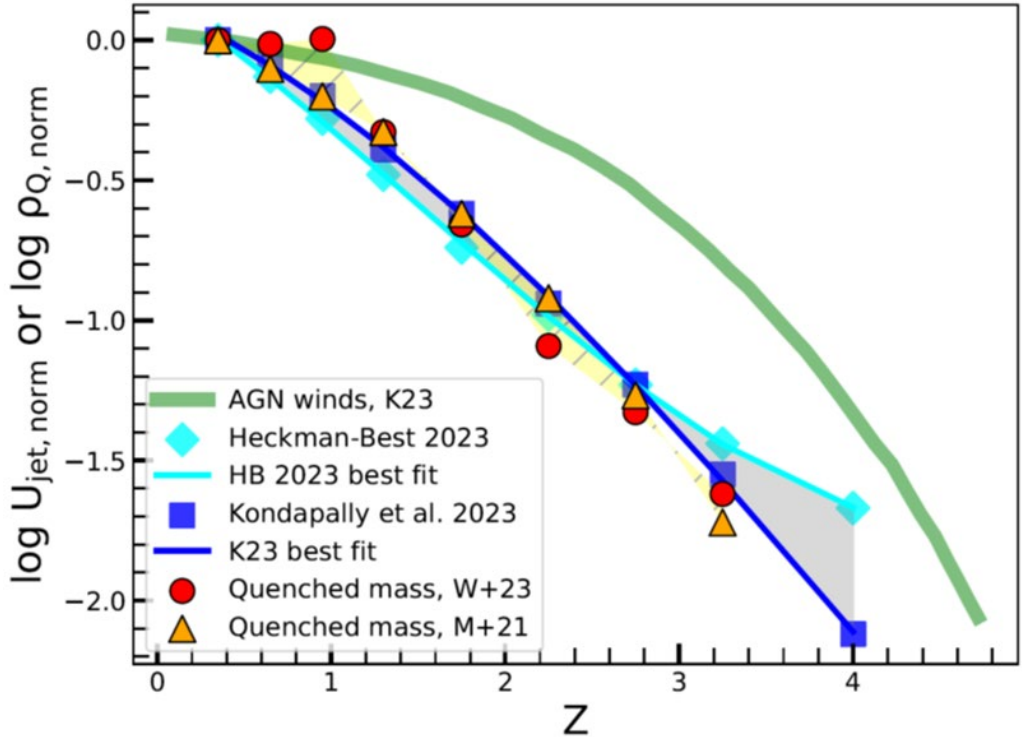

*Figure 1 – We plot the cumulative amount of stellar mass in quiescent massive galaxies, normalized to its value at z = 0.35 as a function of redshift from McLeod et al. (2021- orange) and Weaver et al. (2023 - red). The yellow-hatched-shaded region shows the disagreement between the two measurements. We overplot two independent measurements of the cumulative amount of kinetic energy injected by radio jets, also normalized to its value at z = 0.35 from HB23 (cyan), and K23 (blue). The grey-shaded region shows the disagreements between the two calculations. Note the overall similarity of the evolution of quiescent mass and jet kinetic energy. In green we show the corresponding normalized amount of kinetic energy carried by AGN- or massive star-driven winds (HB23 and K23). This does not match the evolution in mass, and is offset to much earlier times.*

We first take the results from HB23 and K23 to calculate the cumulative amount of injected jet kinetic energy as a function of redshift. We over-plot the results in Figure 1, and normalize the injected energy relative to the cumulative amount at z = 0.35 (see above). The S17 analysis (not shown) yields a result that is entirely consistent with HB23 and K23. Given that HB23, S17, and K23 used different data for the evolution of the radio luminosity functions and different formulae for computing jet kinetic power, the overall agreement between the three calculations is reassuring and provides empirical information about some of the uncertainties in this approach. The three calculations agree to better than 0.1 dex except at z > 3, when less than 3% of the cumulative energy was produced.

### 2.3 Caveats for Jet Energetics

Despite this agreement between HB23, K23, and S17, it is important to consider possible systematic uncertainties in the calculations of jet energetics. As discussed in K23, these relations have a large scatter. While the relations may not be accurate for estimating jet kinetic power for individual sources, we are using the relations to sum over a very large number of radio sources and over a large range in time, which should provide a reasonable total value of kinetic energy. Having said that, there are several possible concerns.

The first point is the degree to which the scaling relations given in Equations 1 and 2 agree with other independent scaling relations in the literature. To test this, K23 computed the production rate of jet kinetic energy using seven different empirical scaling relations between jet kinetic power and radio luminosity. They found that while the normalizations of the different relations varied by about a factor of three, the form of the evolution with redshift was not significantly different. Thus, the evolution of the normalized cumulative kinetic energy shown in Figure 1 should be robust between these different scaling relations.

The next point is that the $P_{jet}$ *vs.* $L_{1.4GHz}$ empirical correlation is based on radio sources over a restricted range in radio luminosity: $L_{1.4GHz} \approx 10^{20}$ to $10^{26}$ W Hz$^{-1}$ (e.g. Heckman & Best 2014). Does the evaluation of the jet power out to z ≈ 3 shown in Figure 1 require a significant extrapolation of this relationship to higher values of $L_{1.4GHz}$? This has also been explored by K23 who find that the peak contribution to jet kinetic energy over this redshift range is from radio sources with $L_{1.4GHz}$ in the range 1 to 5 x $10^{25}$ W Hz$^{-1}$ which lie within the luminosity range calibrated by the X-ray cavities.

Another possible worry is that the X-ray cavity data pertains to the so-called "Low Excitation Radio Galaxies" (LERGs). These are AGN whose energetic output is dominated by the kinetic energy of the radio jets, with little or no radiative output from the supermassive black hole. The other class of radio galaxies called High-Excitation Radio Galaxies ("HERGs") become increasingly important at high radio luminosities (e.g. Best & Heckman 2012). Here, the radiation created by the supermassive black hole (as a QSO) is more energetically important than the radio jet kinetic energy. The HERGs and LERGs may generate different kinds of radio jets and so the radio jet luminosity of HERGs may not agree with equations 1 and 2 above. However, K23 show that LERGs produce 80 to 90% of the volume-averaged $L_{jet}$ over the range z < 2.5.

A related consideration is that Croston et al. (2018) argue that the composition and energetics of jets may be different in radio sources with different morphologies (the FR1 and FR2 sources) and note that the X-ray cavity calibration of jet kinetic power relies on FR1 sources. However, de Jong et al. (2024) show that the relative number of FR1 and FR2 sources does not depend significantly on redshift or radio luminosity above $L_{1.4GHz} = 10^{25}$ W Hz$^{-1}$ and over z = 0.3 – 2.5, so any differences between the FR1 and FR2 sources should not affect the evolution of *normalized* jet kinetic energy.

Finally, one can ask whether the nature of the gaseous environment into which the radio jets propagate is different at high-z from the hot X-ray gas at low-z, or whether the cosmic magnetic fields differ. If so, the efficiency of converting jet kinetic power into radio synchrotron emission may be different (e.g. Hardcastle et al. 2019). The most relevant existing data to evaluate this are observations of radio sources and associated X-ray emission at intermediate redshifts. At low-z, the direct measures of $P_{cav}$ from X-ray data are similar to the corresponding radiative cooling rate measured in X-rays ($L_{cool}$ - Hlavacek-Larrondo et al. 2022), consistent with the idea that heating by jets offsets radiative cooling. At higher redshifts, direct measures of $P_{cav}$ are not possible. However, Hlavacek-Larrondo et al. (2022) and Ruppin et al. (2023) show that values of $P_{cav}$ implied by the observed radio luminosities (see equations 1 and 2) are roughly equal to $L_{cool}$ for radio galaxies at z ≈ 0 to 1.4. This suggests that equations 1 and 2 may not evolve significantly over this redshift range.

We conclude that, at least over the range z ≈ 0 to 2, the assumption that equations 1 or 2 can be used to estimate the jet kinetic energy summed over the radio source population seems reasonable. Figure 1 implies that this epoch encompasses about 90% of the cosmic total of the injected energy by jets.

Finally, we can remake Figure 1 using the cosmic evolution in the radio luminosity function to replace the calculation of the cumulative growth in jet kinetic energy with the purely empirical evolution in the cumulative amount of energy in the form of radio synchrotron emission from radio galaxies. The results are shown in Figure 2 in which we used the radio luminosity functions from Kondapally et al. (2022) and the stellar mass functions for quiescent galaxies from M21. While the agreement above z ≈ 2 is not as good as in Figure 1, there is still a very good correspondence between the two growth curves.

2.4 Comparisons

Comparing the growth in the cumulative mass in quiescent galaxies to the cumulative amount of injected energy in Figure 1 we see that there is excellent agreement. Both are relatively late-time processes, with roughly 90% of the quiescent galaxy mass and jet kinetic energy having been created since z ≈ 2 and about 50% between z = 1 and 2.

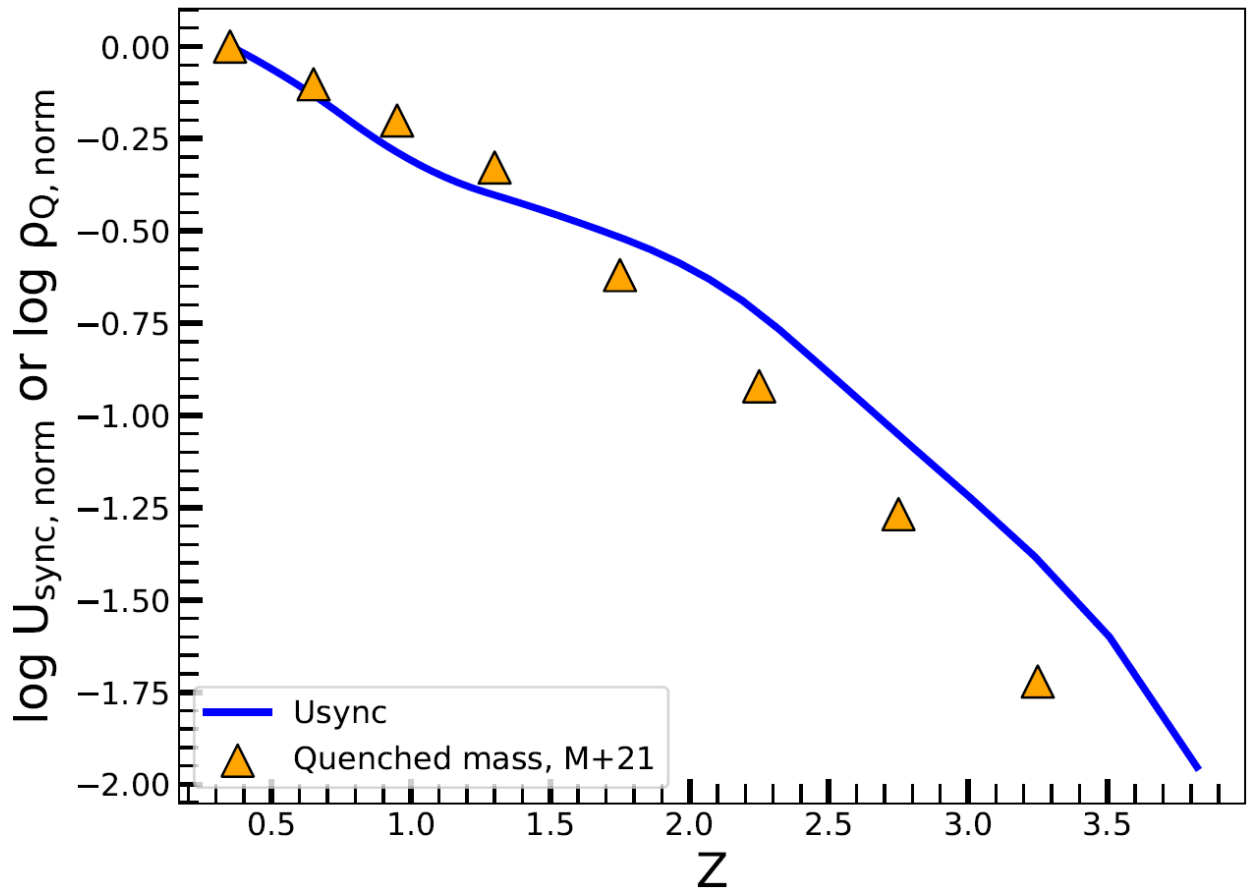

*Figure 2 – As in Figure 1, but replacing the jet kinetic energy with the cumulative amount energy per unit volume in the form of radio synchrotron emission using the radio luminosity functions in Kondapally et al. (2022). The agreement with the growth curve for quench mass (from W21) is not as good as in Figure 1 for z > 2, but there is still good overall synchronization.*

Figure 1 also shows the cumulative growth in the kinetic energy carried by AGN winds rather than jets (from K23 and HB23). This curve is based on the cosmic evolution of the bolometric luminosity production from QSOs (Hopkins et al. 2007) assuming a constant ratio of wind kinetic and bolometric luminosity with cosmic time, based on the data on AGN winds described by HB23.

While the value of this empirical normalization of this ratio is uncertain (HB23, K23), the more important point in Figure 1 is that the evolution of normalized accumulated wind kinetic energy is a poor match to the evolution of accumulated mass in quiescent galaxies. The cosmic evolution of the rate of energy production by accreting supermassive black holes and the rate of star formation closely track one-another (e.g. Heckman & Best 2014). Thus, both the AGN massive-star-driven wind kinetic energies are produced much earlier than is the quiescent mass, reaching about 30% (50%) at redshift 3 (2), when only about 3% (10%) of the quiescent mass has accumulated.

### 2.5 The Evolution of Rate of Major Mergers of Massive Galaxies

We have established a tight relationship between the build-up of mass in quiescent galaxies and the deposition of kinetic energy by radio jets with cosmic time (Figure 1). However, we have not addressed the physical processes that drive this synchronized cosmic evolution.

Any such explanation must account for the fact that the process of quenching involves a transformation in the basic structure of the galaxy. It has been clear for over a decade that quiescent and star-forming galaxies correspond to systems dominated by spheroids and disks respectively, back to at least $z \approx 2$ (e.g. Wuyts et al. 2011; Bell et al. 2012). The only known process for turning disks into spheroids is mergers between two disk galaxies with similar masses. In fact, the analyses of Chiaberge et al. (2015) showed that roughly 90% of radio galaxies at $z \approx 1$ to 2 are associated with recent or ongoing mergers. This was true over a range of roughly five orders-of-magnitude in radio power ($L_{1.4GHz} \approx 10^{23}$ to $10^{28}$ W/Hz).

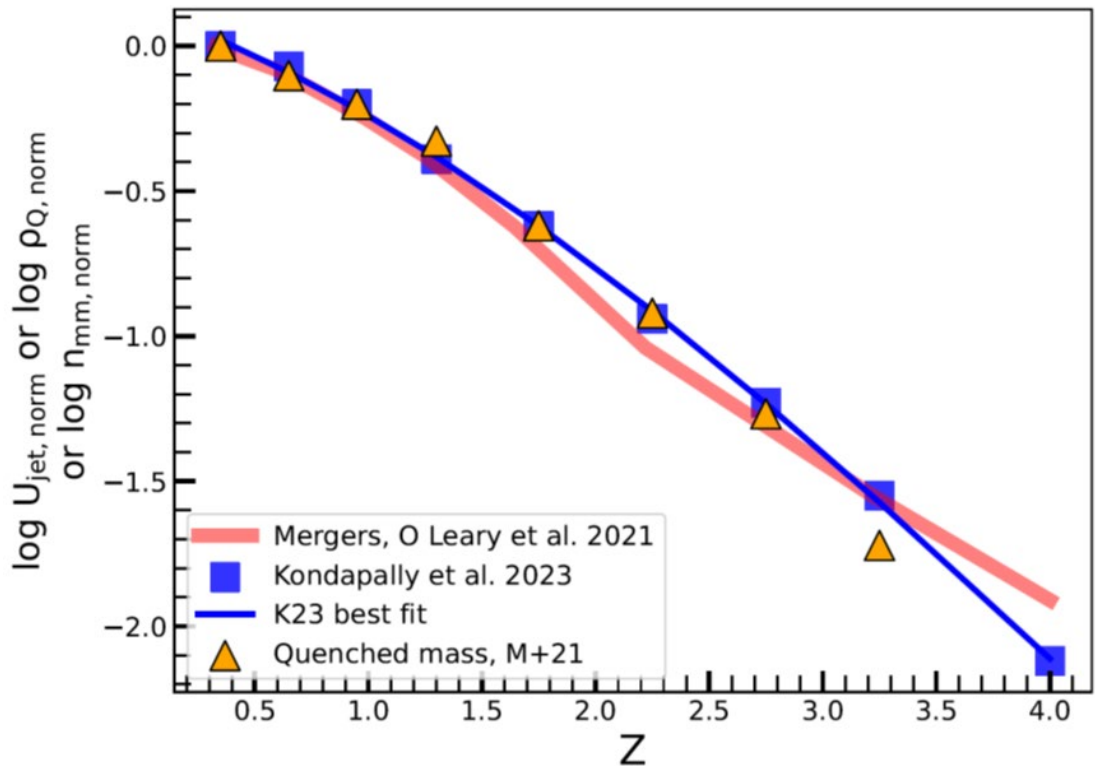

.

*Figure 3 – The cumulative growth in the number of major mergers of massive galaxies per unit volume, (see text) normalized to the value at z = 0.35 as a function of redshift from McLeod et al. (2021- red). This*

*is compared to measurements of the normalized cumulative amount of kinetic energy per unit volume injected by radio jets (K23- blue) and the normalized mass per unit volume in quiescent galaxies per unit volume M21 – orange). All three curves agree extremely well with one another. This suggests that major mergers of massive galaxies trigger the production of jets, which then quench star formation and also transform the galaxy structures from disks to spheroids*

To further evaluate the role of major mergers in driving the evolution of the radio jets and quiescent galaxies seen in Figure 1, we have used the results in O'Leary et al. (2021). They computed galaxy merger rates as a function of redshift in bins of the stellar mass of the resulting galaxies and the stellar mass-ratio between the two progenitors. From this, we have selected the computation for massive descendants ($M_* \geq 10^{11}$ $M_\odot$), as appropriate for radio galaxies (e.g. Best et al 2023), and for major mergers (mass ratio < 4). We will refer to these as major mergers of massive galaxies.

As in the case of jet kinetic energy and quiescent galaxy mass, we use the results in O'Leary et al. to compute the cumulative number of major mergers of massive galaxies per co-moving volume element as a function of redshift, normalized to the value at z = 0.35. The result is shown in Figure 3 compared to the K23 data for jet kinetic energy and the M21 data for quiescent galaxies. We see a remarkable agreement between the cosmic evolution of major mergers of massive galaxies, and both the amount of kinetic energy carried by radio jets, and the mass created in quiescent galaxies. It is hard not to conclude that major mergers of massive galaxies drive both radio jet production and the creation of spheroid-dominated galaxies.

2.6 The Sequencing of Events

We have established tight relationships between the build-up over cosmic time of major mergers of massive galaxies, of the stellar mass in quiescent galaxies, and of the deposition of kinetic energy by radio jets with cosmic time (Figures 1 and 3). However, this does not directly demonstrate the nature of the causal connection between these three phenomena. We believe that the most natural interpretation is that a major merger of star-forming disk galaxies both triggers an outflow of kinetic energy from radio jets that quench star-formation, and transforms the disks to a spheroid-dominated galaxy. However, the causal connection could be different.

For example, one popular scenario (e.g. Sanders et al. 1988; Hopkins et al. 2006) is that a major merger leads to a powerful starburst and AGN and that the winds they generate sweep out the remaining gas, quenching star formation. The disks merge into an elliptical galaxy and after some time elapses, a hot halo of gas is created, which cools and fuels radio jets (e.g. Best et al. 2014). Is this case the quenching leads to the radio jets rather than the other way around. How can we then determine the correct direction of the causal connection?

One way to assess the direction of causality is to consider star-formation in radio galaxies. In the low-z universe, powerful radio jets are almost all found in massive elliptical galaxies that have long-since been quenched (e.g. Best & Heckman 2012). In these cases, the radio jets are acting to prevent further star-formation by heating the X-ray gas that would otherwise cool and form stars (e.g. McNamara & Nulsen 2007). This would be consistent with the "quenching precedes the radio jets" scenario. For radio jets to precede (and cause) quenching, they must be produced in star-forming galaxies.

In fact, Kondapally et al. (2022) have shown that the fraction of radio galaxies that are star-forming rather than quiescent rises with redshift, and that they become the majority of the population at z > 0.9 (see also S17 and Das et al. 2024). These galaxies lie on or near the star-forming main sequence, so most radio galaxies during the main "quenching era" (z ≈ 1 to 2, see Figure 1) were forming stars at the time the jets were created. This is consistent with the scenario in which radio jets can later quench star-formation, and then "maintain" a low-level of subsequent star-formation in quiescent massive galaxies.

Another argument against the merger/quenching/jet-production scenario is that the quenching associated with winds driven by a starburst or AGN should follow the "AGN Wind = Massive-Star Wind" cumulative growth rate curve shown in Figure 1 (see above). This is clearly not the case. Moreover, the great majority of radio galaxies contributing to Figure 1 show a sign of neither a powerful QSO-like AGN or starburst (K23, Kondapally et al. 2022).

Taken together, the arguments above support the conclusion that the dominant causal connection appears to be one in which mergers lead to radio jets which lead to the quenching of star-formation in massive galaxies.

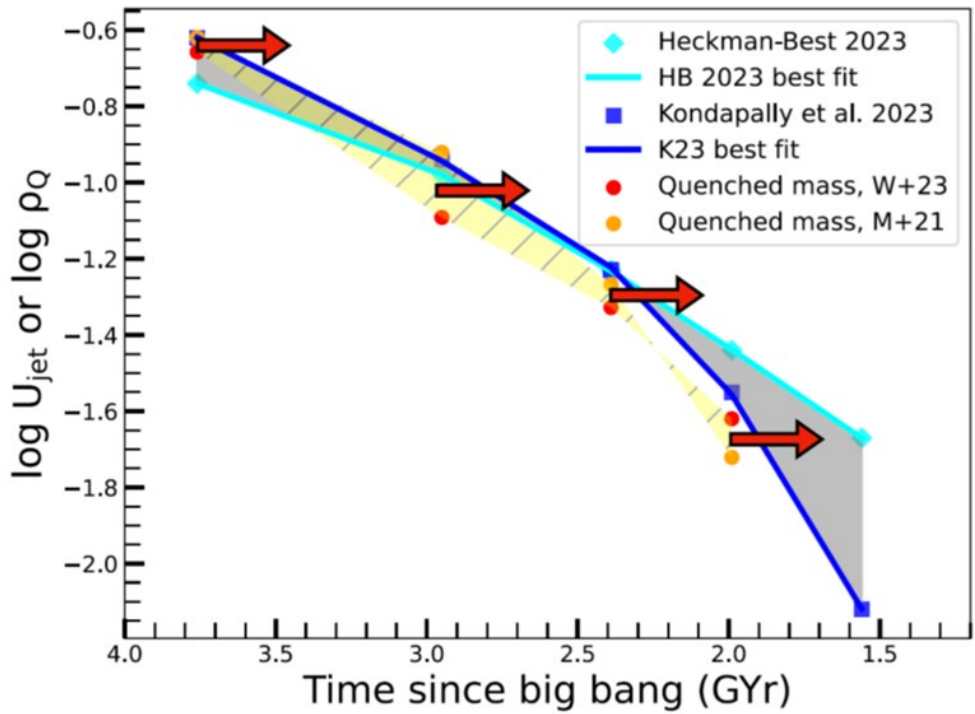

*Figure 4 – We replot the same data as in Figure 1, but now replace redshift with the time since the Big Bang on the x-axis, and zoom-in on the epoch before z = 1.5 (t < 4 Gyr). The rightward facing arrows*

*show the effect of a time-lag of 0.3 Gyr between the time star-formation is actually quenched and when stellar evolution results in the red spectral energy distribution of a quiescent galaxy (see text).*

One additional consideration for quenching by jets is that there is a built-in timescale set by stellar evolution that will introduce a lag between when star-formation is shut-off by radio jet feedback (quenching occurs), and when stellar evolution leads to the red spectral energy distribution that is used to categorize quiescent galaxies. Observations of quiescent massive galaxies at z ≈ 1.5 to 2.5 (Belli et al. 2019; Park et al. 2023; Park et al. 2024) yield timescales ranging from less than 100 Myr to over a Gyr after quenching to produce the observed galaxy colors and spectra. In Figure 4 we show a zoom-in on the portion of Figure 1 at z ≈ 1.5 to 3.3. We replot this using the time since the Big Bang on the x-axis rather than redshift. We indicate with right-facing arrows the time at which star-formation must have ceased in the past to produce a quiescent galaxy 0.3 Gyr later (a mid-range value from the literature cited above).

This figure shows that the jet-induced quenching model would require most massive galaxies to quickly (≤ 0.3 Gyr) evolve from star-forming to quiescent at these redshifts. This need not be the case at low-z, since a time-lag of 1 Gyr would (for example) only correspond to an offset of Δz = 0.17 at z = 0.65. This could be accommodated by the results in Figure 1. Indeed, there is evidence that rapid quenching becomes more important with increasing redshift (e.g. Belli et al. 2019).

## 3. THE ENERGETICS OF QUENCHING

In this section, we consider whether the amount of kinetic energy carried by radio jets would be sufficient to quench star formation (as well as just to subsequently maintain the quenched state). A detailed answer to this question is not possible because the precise mechanisms by which jet energy couples to the gas and the specific effects this has are matters of on-going debate (e.g. Wagner & Bicknell 2011; Mukherjee et al. 2016).

Here we ask a simpler question: how does the jet kinetic energy compare to the gravitational binding energy of the baryons in the galaxy? To do this, we begin by considering the normalizations of the kinetic energy and quenched mass shown in Figure 1. HB23 calculated a total amount of accumulated (i.e., integrated across redshift) jet kinetic energy per unit volume of $U_{jet}$ = 2.7 x $10^{57}$ erg Mpc$^{-3}$, while K23 and S17 found about half this amount. The most detailed analysis of the stellar mass per co-moving volume element for quiescent galaxies in the present-day universe comes from the analysis of Weigel et al. (2016) based on the SDSS. This yields a value of $\rho_Q$ = 1.2 x $10^8$ M$_\odot$ Mpc$^{-3}$.

Let us now consider the mass-dependence of these normalizations. Janssen et al. (2012) and Kondapally et al. (2022) showed that the probability of a galaxy creating a radio source with a given luminosity (given jet mechanical luminosity) was proportional to $M_*^{2.5}$ for LERGs and $M_*^{1.5}$ for HERGs from z ≈ 0 to at least 1.5. As noted in section 2.3, K23 showed that LERGs produce 80 to 90% of the volume-averaged $L_{jet}$ over the range z < 2.5. Adopting the LERG scaling would then imply that the total amount of jet kinetic energy delivered to the environs of the galaxy is

$KE_{jet} \propto M_*^{2.5}$. The gravitational binding energy of a galaxy $E_{bind} \propto M_* v_{circ}^2$, where $v_{circ}$ is the galaxy circular velocity. Both the Tully-Fisher and Faber-Jackson relations give $v_{circ} \propto M_*^{0.25}$ (Cimatti et al. 2020). This then yields $KE_{jet}/E_{bind} \propto M_*$. Thus, jet feedback should be more important in more massive galaxies (the opposite to the case of feedback from massive stars).

This can be further quantified by using the value of $U_{jet}$ integrated over the stellar mass function of quiescent galaxies and weighting $KE_{jet}$ by $M_*^{2.5}$. We can quantify the cumulative average amount of $KE_{jet}$ delivered to a quiescent galaxy as a function of its stellar mass. This yields a value $2.6 \times 10^{60}$ $(M_*/10^{11} M_\odot)^{2.5}$ ergs for HB23 and half that amount for K23 and S17. It is also important to note that integrating the relationship $KE_{jet} \propto M_*^{2.5}$ over the stellar mass function of quiescent galaxy will diverge at large $M_*$ unless the mass function has sharp cut-off at the high-mass end (like the exponential cut-off in the Schechter function).

Noting that the jet will affect the galaxy's gas, and defining the ratio of gas to stellar mass as $f_{gas}$, this can also be written as $E_{jet}/M_* = 13.0\ (6.5) \times 10^{15} f_{gas}^{-1} (M_*/10^{11} M_\odot)^{1.5}$ erg/gm for HB23(K23,S17). We can use this to define an equivalent velocity $v_{eq} = (2\ E_{jet}/f_{gas}\ M_*)^{1/2} = 1610\ (1140)\ f_{gas}^{-0.5} (M^*/10^{11} M_\odot)^{0.75}$ km/s for HB23 (K23,S17). This exceeds the escape velocity of massive galaxies, implying that the jet kinetic energy can have a powerful impact on gas in the pgalaxy. On larger scales (out to the virial radius) the stellar-mass/halo-mass relation (Girelli et al 2020) implies $f_{gas} \approx 3$ to 4 for $M_{halo} \approx 10^{12} M_\odot$ to $10^{13} M_\odot$ (including stars in both the central and satellite galaxies – Mitchell & Schaye 2022). Thus, jets can have a significant impact on the halo gas (see Donahue & Voit 2022).

An interesting additional point is to note that Figure 1 implies that the ratio of $U_{jet}/\rho_Q$ is independent of redshift. In the context of an ensemble of individual quiescent galaxies, the average ratio of $<E_{jet}/M_*>$ can therefore depend only on $M_*$ (as $M_*^{1.5}$ – see above), but not on redshift. This has very interesting implications, especially for understanding what physical processes specify the ratio of $E_{jet}/M_*$ following a major merger of massive disk galaxies.

As a final exercise ("sanity check") we can compare our results to the measurement of the thermal Sunyaev-Zel'dovich effect in a stacked sample of massive galaxies at $z \approx 1$ (Meinke et al. 2023). The mean stellar mass of the galaxies in the stack is $2.5 \times 10^{11} M_\odot$. The ratio of jet kinetic energy to stellar mass accumulated by $z \approx 1$ would be 2.6 (HB23) or 1.3(K23) $\times 10^{16}$ erg/gm at $M^* = 2.5 \times 10^{11} M_\odot$, implying $KE_{jet} = 1.3 \times 10^{61}$ (HB23) or $6.5 \times 10^{60}$ (K23,S17) ergs for these very massive galaxies. These values can be compared to the mean amount of thermal energy found by Meinke et al. (2023) of $8 \times 10^{60}$ ergs. This is reassuring agreement. In the same vein, HB23 pointed out that the cumulative amount of jet energy at $z \approx 0$ is very close to the amount needed to explain the thermal and physical properties of hot gas in groups and clusters, like the $L_X$-$T_X$ relation and the baryonic mass fractions as a function of halo mass (Donahue & Voit 2022).

4. DISCUSSION & FUTURE DIRECTIONS

The remarkable synchronization of the build-up of stellar mass in quiescent galaxies and the transport of kinetic energy by radio jets over cosmic time (Figure 1) is strong evidence that radio jets are responsible for quenching star-formation in massive galaxies. Moreover, simple energetic considerations establish that radio jets can – in principle – have drastic effects on the gas-phase baryons in and surrounding massive galaxies. We have seen that this synchronized cosmic evolution is plausibly driven by the corresponding evolution in the rate of major mergers of massive galaxies (Figure 3). Still, this evidence must be regarded as circumstantial until we understand the physical processes by which the major mergers trigger jets and jets quench star formation.

As described above, direct evidence for the feedback effect of jets consists of observations of cavities in the hot X-ray gas that have been inflated by radio jets (McNamara & Nulsen 2007 and references therein). However, these observations have only been made at low redshift, and the effect of the feedback is to inhibit the cooling of hot gas that would otherwise be accreted and form stars. These galaxies were quenched long ago, so the feedback is maintaining the low-level of star formation, not transforming the galaxy from star-forming to quiescent.

Thus, there appear to be two modes of jet feedback in massive galaxies: one that transforms galaxies by quenching star-formation and one that inhibits star-formation in quenched galaxies. Quenching is often envisaged as resulting from the ejection of dense gas from galaxies. However, a plausible alternative is that the feedback quenches galaxies by heating, lifting, and/or disrupting the gas in the circum-galactic medium that would otherwise be accreted and fuel future star formation (e.g. Donahue & Voit 2022). In this case, the physical processes and locations of the maintenance and quenching modes of jet feedback may be qualitatively similar.

Figures 1 and 3 suggest that the best place to observe the quenching mode in operation would be in radio galaxies at $z \approx 1$ to 2 when roughly 50% of the quenching occurred. To date, direct observations of jet feedback beyond the local universe have been at higher redshifts than this, typically studying small samples of more extreme radio galaxies. Integral field unit spectra of the rest-frame optical emission-lines in these radio galaxies show clear evidence that the warm ionized component of the interstellar/circum-galactic medium is being accelerated to high velocities by the jets. Currently, these are primarily ground-based data with limited spatial resolution (e.g. Nesvadba et al, 2017). With JWST this is just now being revisited with both high spatial resolution and complete coverage of the main diagnostic emission-lines (Roy et al. 2024; Saxena et al. 2024; Wang et al. 2024). The main needs are to build up a sample large enough to assess how the feedback energetics depend on the jet power, obtain data on radio galaxies in the "quenching era", and seek evidence that star-formation is in fact being terminated in these galaxies.

ACKNOWLEDGEMENTS

We thank the referee Mark Voit for a thought-provoking report that helped us improve the paper. TH also thanks the Aspen Center for Physics for hosting the meeting "Toward a Holistic

Understanding of the Multi-scale, Multiphase Circumgalactic Medium" where useful discussions with colleagues helped inform this paper.